\documentstyle[preprint,aps]{revtex}
\begin{document}
\draft
\preprint{\vbox{\hbox{CLNS 97/1462}\hbox{SISSA 15/97/FM}}}
\title{Comment on ``A new efficient method for calculating perturbative
energies using functions which are not square integrable'':
regularization and justification}
\author{C.K. Au}
\address{Department of Physics and Astronomy, University of South
Carolina,
Columbia, SC 29208.}
\author{Chi--Keung Chow}
\address{Newman Laboratory of Nuclear Studies, Cornell University,
Ithaca,
NY 14853.}
\author{Chong--Sun Chu}
\address{SISSA, Via Beirut 4, 34013 Trieste, Italy.}
\date{\today}
\maketitle
\begin{abstract}
The method recently proposed by Sk\'ala and \v C\'i\v zek for
calculating
perturbation energies in a strict sense is ambiguous because it is
expressed as a ratio of two quantities which are separately divergent.
Even though this ratio comes out finite and gives the correct
perturbation
energies, the calculational process must be regularized to be justified.
We examine one possible method of regularization and show that the
proposed method gives traditional quantum mechanics results.
\end{abstract}
\narrowtext
Recently, in a letter in this journal \cite{SC}, Sk\'ala and \v C\'i\v
zek
(SC) proposed a method to calculate perturbation energies using
non-sqaure-integrable functions.
The method of SC is further augmented in a comment by Guardiola and Ros
(GR) \cite{GR}.
The purpose of our present comment is to further point out that in a
strict
sense, the SC method for the perturbation energies can result in a ratio
of
two divergent quantities, so that while this ratio may remain finite in
a
practical numerical calculation, a regularization procedure is needed to
justify the finite result.
We have examined one such possible regularization procedure and made
contact
between the SC method and traditional quantum mechanics (QM) results.

Briefly, the SC method regards the $n$-th order perturbation equation as
a
parametric differential equation with $E_n$ as the parameter,
\begin{equation}
(H_0 - E_0) \psi_n (E_n,x) = (E_n - \tilde V_n) \psi_0 (x),
\label{master}
\end{equation}
where
\begin{equation}
H_0= -\frac{d^2}{dx^2} +V_0,
\end{equation}
and
\begin{equation}
\tilde V_n \psi_0 \equiv V_1 \psi_{n-1} - \sum_{i=1}^{n-1} E_i
\psi_{n-i}.
\label{defV}
\end{equation}
Here we have chosen to introduce $\tilde V_n$ as the effective
perturbation in
the $n$-th order equation.
For $n=1$, the sum in R.H.S.~of Eq.~(\ref{defV}) vanishes and $\tilde
V_1$
is the same as the real perturbation $V_1$.
We also adopt the convention that
all the wavefunctions are physical unless the dependence on the
parameter $E_n$ is explicitly displayed.
For this form of Eq.~(\ref{master}), the $n$-th order equation is
similar in
form to the first order equation.
$\tilde V_n$ is a known function since all lower order quantities are
assumed
known.
In traditional QM perturbation theory, one left multiplies
Eq.~(\ref{master})
with the zeroth order wave function $\psi_0$.
Under the usual normalization conditions,
\begin{equation}
\langle \psi_0|\psi_0 \rangle = 1 \qquad \hbox{and} \qquad
\langle \psi_0|\psi_i \rangle = 0 \quad \forall i \neq 0,
\label{norm}
\end{equation}
one recovers the standard QM result
\begin{equation}
E_n = \langle \psi_0 |\tilde V_n| \psi_0 \rangle
=  \langle \psi_0 |V_1| \psi_{n-1} \rangle.
\end{equation}
Once $E_n$ is correctly obtained, $\psi_n(E_n, x)$ is obtained by
solving the
ordinary differential equation in Eq.~(\ref{master}).

SC propose that instead of obtaining $E_n$ in the standard way first,
one
treats Eq.~(\ref{master}) as a parametric ordinary differential equation
with
$E_n$ a parameter, and go on to show that
\begin{equation}
E_n = {- \psi_n(0, x_0)\over \psi_n (1, x_0) - \psi_n (0, x_0)} \equiv
{- \psi_n(0, x_0)\over F(x_0)},
\label{para}
\end{equation}
where $x_0$ is a point such that the boundary conditions
\begin{equation}
\psi_n(E_n,x_0) = 0, \qquad n = 0,1, \dots
\label{bc}
\end{equation}
are met for the physicsal energies $E_n$.
Since neither 0 nor 1 is the necessary $n$-th order energy correction
$E_n$,
the functions $\psi_n(0, x)$ and $\psi_n(1, x)$ are in general {\it
not\/}
square integrable and hence the name for the method.

Since the ground state wave function vanishes only at the end points of
the boundary\footnote{For purpose of illustration, we consider QM on a
half
line $[0, \infty)$.} and
the nodal points of the wave functions of the excited states shift upon
turning on the perturbation, the {\it only\/} choice for $x_0$
consistent
with the boundary conditions (\ref{bc}) is $x_0=\infty$.
In a practical numerical calculation, which is always carried out in
between finite ranges, $x_0$ is assigned an arbitrarily large but finite
value. But  as $x_0$ approaches infinity, both $\psi_n(0, x_0)$ and
$\psi_n(1, x_0)$ diverge, and a regularization process is needed to make
sense of Eq.~(\ref{para}).

In the form of Eq.~(\ref{master}), taking into account of the advantage
of
its similarity in form to the first order equation, $\psi_n(\alpha, x)$
can
easily be solved, say, using the Dalgarno--Lewis method \cite{DL} or
logarithmic perturbation method \cite{AA,Tu} to obtain,
\begin{equation}
\psi_n(\alpha,x) = - \psi_0(x) \int_b^x dy\, {1\over \psi_0^2(y)}
\int_a^y dz
\,(\alpha - \tilde V_n) \psi_0^2(z) ,
\label{sol}
\end{equation}
where $a$ and $b$ are appropriate constants to satisfy the boundary
conditions,
in agreement with the results of GR.

{}From Eq.~(\ref{sol}), one recovers the universal functions $F(x)$
(given as
Eq.~(15) in Ref.~\cite{GR})
\begin{equation}
F(x) = - \psi_0(x)
\int_b^x dy \,{1\over \psi_0^2(y)} \int_a^y dz \,V(z) \psi_0^2(z).
\end{equation}
Together with Eq.~(\ref{para}), one sees that the $n$-th order
perturbation
energy $E_n$ is given by
\begin{equation}
E_n = {J(\tilde V_n, x_0)\over J(1, x_0)},
\label{En}
\end{equation}
where the functional $J(V,x)$ is given by
\begin{equation}
J(V,x) \equiv \int_0^x dy \,{1\over \psi_0^2(y)} \int_0^y dz \,V(z)
\psi_0^2(z).
\label{J}
\end{equation}
and the boundary condition at the endpoints has been taken care of
appropriately.

Next, we would like to point out that at least in the example of the
ground
state of the $x^4$ anharmonic oscillator, the expansion of $E_n$ in
Eq.~(\ref{En}) can be ill-defined because both the numerator and the
denominator diverge as $x_0 \to\infty$.
This can be easily seen by combining the well-known results that $\tilde
V_n$
is of polynomial form in the Bender--Wu \cite{BW} $x^4$ anharmonic
oscillator
and the mean value theorem.

{}From the form of Eq.~(\ref{J}), one does {\it not\/} expect {\it a
priori\/}
that in the limit $x_0 \to \infty$, the ratio $J(\tilde V_n, x_0)/J(1,
x_0)$
becomes finite and $x_0$ independent even though numerically this comes
out
to be so.
Hence to make sense out of Eqs.~(\ref{En}) and (\ref{J}), a
regularization
procedure is in order.
One can justify the numerical result obtained by assigning an
arbitrarily
large but finite value to $x_0$ only after the result is regularized and
the
limit is proven to exist.

The regularization procedure being proposed here is similar to the one
we
previously used in the extension of logarithmic perturbation theory to
excited bound states in one dimension by appropriately mixing in the
ghost
state \cite{We}.
\begin{mathletters}
For the zeroth order solution (unperturbed state), instead of using the
square integrable eigenstate wave function $\psi_0$, we can mix in the
non-square-integrable ghost state $\chi_0$ by defining
\begin{equation}
\Psi_0(x) \equiv \psi_0(x) + i\sigma\chi_0(x),
\end{equation}
\begin{equation}
\rho(x) \equiv \Psi_0^2(x).
\label{rho}
\end{equation}
and
\begin{equation}
J_\sigma[S] \equiv \int_0^\infty dy \,{1\over\rho(y)} \int_0^y dz \,
\rho(z)
S(z).
\end{equation}
\end{mathletters}
Note that in Eq.~(\ref{rho}), $\rho(x)$ is the ordinary square of
$\Psi_0(x)$,
not $|\Psi_0(x)|^2$.
Then Eq.~(\ref{En}) can be rewritten on firm mathematical ground as
\begin{equation}
E_n = \lim_{\sigma\to0}{J_\sigma[\tilde V_n]\over J_\sigma[1]}
\label{En2}
\end{equation}
(\ref{J})
Now, we can show that the limit in Eq.~(\ref{En2}) is the
well-defined.
This follows from \cite{We}
\begin{eqnarray}
J_\sigma[S] &=& {i\over\sigma}
\int_0^\infty dy \,\Psi_0(y) \psi_0(y) S(y) \nonumber\\
&=& {i\over\sigma} \int_0^\infty dy \,\psi_0^2(y) S(y) + \cdots,
\label{En3}
\end{eqnarray}
where $\cdots$ is a  $\sigma$ independent term.
Upon substituting Eq.~(\ref{En3}) into Eq.~(\ref{En2}), we recover
\begin{equation}
E_n={\int_0^\infty dy \,\psi_0^2(y) \tilde V_n(y) \over
\int_0^\infty dy \,\psi_0^2(y)}
= \int_0^\infty dy \,\psi_0(y) V_1(y) \psi_{n-1}(y)
\end{equation}
which is the ordinary QM result upon using Eq.~(\ref{defV}) and
(\ref{norm}).
Hence, we have provided a rigorous justification of the SC method.
It is interesting to note that we have also utilized
non-square-integrable
functions through the ghost state mixing.

To sum up, we see that the SC method correctly gives the perturbation
energies, but as a ratio of two divergent quantities.
We have regularized it through ghost state mixing and our final result
is
independent of the mixing parameter $\sigma$.
It is only after establishing the existence of the limit in
Eq.~(\ref{En2})
that we can accept the numerical convergence in Eq.~(\ref{En}) advocated
in
the SC method.

\acknowledgments
The work of C.K.C. is supported in part by the National Science
Foundation.

\end{document}